\documentstyle[12pt]{article}
\title{Conditional symmetry and spectrum of the one-dimensional
  Schr\"odinger equation}  
\author{R.Z.~Zhdanov \\ \small Arnold-Sommerfeld Institute for
  Mathematical Physics,\\ \small Leibnitzstra\ss e 10, 38678
  Clausthal-Zellerfeld, Germany\\ \small Electronic mail {\tt
    asrz@pt.tu-clausthal.de}
\thanks {On leave from the Institute of Mathematics of the Academy of
  Sciences of Ukraine, Tereshchenkivska Str.3, 252004 Kiev, Ukraine}}
\date{}
\let\p\partial 
\let\ve\varepsilon 
\let\ds\displaystyle
\newtheorem{theo}{Theorem}
\begin{document}
\maketitle
\begin{abstract}
  We develop an algebraic approach to studying the spectral properties
  of the stationary Schr\"odinger equation in one dimension based on
  its high order conditional symmetries. This approach makes it
  possible to obtain in explicit form representations of the
  Schr\"odinger operator by $n\times n$ matrices for any $n\in {\bf
    N}$ and, thus, to reduce a spectral problem to a purely algebraic
  one of finding eigenvalues of constant $n\times n$ matrices.  The
  connection to so called quasi exactly solvable models is discussed.
  It is established, in particular, that the case, when conditional
  symmetries reduce to high order Lie symmetries, corresponds to
  exactly solvable Schr\"odinger equations. A symmetry classification
  of Sch\"odinger equation admitting non-trivial high order Lie
  symmetries is carried out, which yields a hierarchy of exactly
  solvable Schr\"odinger equations. Exact solutions of these are
  constructed in explicit form. Possible applications of the
  technique developed to multi-dimensional linear and one-dimensional
  nonlinear Schr\"odinger equations is briefly discussed.
\end{abstract} 

\section*{I.\ Introduction}
Basic motivation for introducing conditional symmetries (the term \lq
conditional symmetry\rq\ was suggested for the first time by {\sc
  Fushchych} \cite{fu1}--\cite{fu3}) was a necessity to find a
symmetry background of a quickly growing variety of exact solutions of
nonlinear partial differential equations that could not be obtained
within the framework of the classical Lie approach. An intensive
search of such solutions was begun independently and almost
simultaneously by {\sc Fushchych} with collaborators (see,
\cite{fu4,fu5} and references therein), {\sc Clarkson} \& {\sc
  Kruskal} (\lq the direct reduction method\rq\ \cite{cla}), {\sc
  Olver} \& {\sc Rosenau} (\lq non-classical reduction\rq\ 
\cite{olv1}) and {\sc Winternitz} \& {\sc Levi} \cite{win}. A number
of examples of nonlinear partial differential equations in two, three
and even four dimensions having non-trivial conditional symmetries is
growing rapidly. In particular, it has been established by {\sc
  Fushchych, Zhdanov and Revenko} \cite{fu6}--\cite{zh2} that such
fundamental equations of the modern quantum field theory as the
four-dimensional nonlinear d'Alembert, Dirac, Levi-Leblond, Maxwell
and Yang-Mills equations possess {\em infinite} conditional symmetries,
while their Lie symmetries are finite only.

On the other hand, much less attention is devoted to the study of
conditional symmetries of linear differential equations (though the
first example of conditional symmetry has been obtained by {\sc
  Bluman} and {\sc Cole} for the one-dimensional linear heat equation
\cite{blu}). In view of the role played by conditional symmetries in
the theory of nonlinear differential equations one can expect that
application of these to linear equations will also be rich in
results. In the present paper we establish the rather unexpected (at
least for the author) fact that conditional symmetries can be
effectively applied to study spectral properties of the stationary
Schr\"odinger equation
\begin{equation}
  \label{0.1}
  \psi_{xx}=(\ve+V(x))\psi.
\end{equation}

In particular, we will prove that it is conditional symmetry that
is responsible for a phenomenon of so called \lq quasi exact
solvability\rq\ of some specific class of equations (\ref{0.1})
\cite{tur}--\cite{ush}. And what is more, the case when conditional
symmetries are equivalent to Lie symmetries will be shown to yield
exactly solvable Schr\"odinger equations.

All principal approaches to a construction of (quasi) exactly solvable
models (apart from the specific ways of an implementation of these)
are based on a possibility to construct a basis $u_1(x)$,\ $u_2(x)$,\ 
$\ldots$,$u_n(x)$ of some invariant space ${\cal V}_n$ of the
Schr\"odinger operator $S=\p_x^2-V(x)$. This means that there should
exist constant $n\times n$ matrix $\|\Lambda_{jk}\|$ such that the
following conditions are fulfilled:

\begin{equation}
  \label{0.2}
  S u_j(x)\equiv
  (\p_x^2-V(x))u_j(x)=\sum\limits_{k=1}^n\Lambda_{jk}u_k(x),\quad 
  j=1,2,\ldots,n.
\end{equation}

Given such functions $u_j(x)$, a procedure for calculating the spectrum
of the Schr\"odinger operators (or, more precisely, a part of the
spectrum) is completely algebraic.  Let $\vec a^{j}=(a_1^{j}, a_2^{j},
\ldots, a_n^{j}),\ j=1,\ldots,m,\ m\le n$ be a complete system of
eigenvectors of the $(n \times n)$ matrix $\Lambda
=\|\Lambda_{jk}\|_{j,k=1}^n$ and $\lambda_1,\ldots,\lambda_m$ be their
eigenvalues, namely
\begin{equation}
  \label{0.3}
  \sum\limits_{j=1}^n\Lambda_{jk}a^{l}_j=\lambda_la^{l}_k,\quad
  j=1,2,\ldots,m. 
\end{equation}

Then, the function
\begin{equation}
  \label{0.4}
  \psi_{k}(x)=\sum\limits_{j=1}^na^k_ju_j(x)
\end{equation}
is easily seen to satisfy the equation (\ref{0.1}) with
$\ve=\lambda_k$ under arbitrary $k=1,2,\ldots,m$.

Saying it another way, after being restricted to a linear space ${\cal
  V}_n$ with basis functions $u_1(x)$,\ $u_2(x)$,\ $\ldots$,$u_n(x)$
the Schr\"odinger operator becomes a matrix operator.  Thus, a
reduction of a differential operator to a matrix operator takes place.
But such a procedure is quite a common routine in the theory of Lie
symmetries of differential equations. Indeed, if we restrict a partial
differential equation having $N$ independent variables to a subset of
its solutions invariant under a one-parameter subgroup of the Lie
group admitted by the equation in question, then it is reduced to a
partial differential equation with $N-1$ independent variables.  Such
a procedure is called symmetry reduction of differential equations
(for more detail, see e.g. \cite{fu4,olv2,ovs}). Taking $N=1$ (the
case of ordinary differential equation) we obtain as a reduced
equation a differential equation with $N=0$, i.e.\ an algebraic
equation!

One of the main aims of the present paper is to show that the idea of
symmetry reduction, when formulated in an appropriate way, can be
applied effectively to an algebraization of the problem of describing
spectrum of the Schr\"odinger operator.

As mentioned above classical Lie symmetries of partial differential
equation do not give all possible reductions. More general symmetries
responsible for a possibility of reducing the order of differential
equations are conditional symmetries.  Roughly speaking, the necessary
and sufficient condition providing a possibility to reduce a number of
variables in a given partial differential equation is a requirement of
conditional invariance (\cite{fu9}). It will be established that a
similar situation takes place for the Schr\"odinger equation
(\ref{0.1}).  Symmetries providing reducibility of differential
equation (\ref{0.1}) to a system of algebraic equations of the form
(\ref{0.2}) are exactly the high order conditional symmetries
introduced independently by {\sc Zhdanov} \& {\sc Fushchych}
\cite{fu10,zh3} and {\sc Fokas} \& {\sc Liu} \cite{fok} (see also
\cite{nik2}--\cite{kap}).
 
It should be emphasized that considerations of the present paper are
purely {\em algebraic}. The method of conditional symmetries making it
possible to study spectral properties of the Schr\"odinger operator
$S$ gives no information about analytical properties of the
corresponding eigenfunctions. In each specific case such properties of
the eigenfunctions obtained as square integrability, asymptotic
behavior, singularities etc. should be studied independently (see,
e.g.  \cite{gon}). The reason is that this method (and
group-theoretical, symmetry methods in general) exploits algebraic
properties of the solution set of equation (\ref{0.1}) (or its part)
as a whole and, roughly speaking, is independent of analytical
properties of specific solutions.

\section*{II.\ Conditional symmetry of the Schr\"odinger equation}

Consider the $n$-th order differential operator
\begin{equation}
\label{1.1}
  Q=\sum\limits_{j=0}^{n}q_{j}(x)\partial^j_x,
\end{equation}
where $\partial^0_x=1,\ \partial^{j+1}_x={d\over dx}\, \partial^j_x$
and functions $q_j(x)$ are supposed to be independent of $\ve$.

Following \cite{fu11,zh5} we say that equation (\ref{0.1}) is
conditionally invariant with respect to the operator $Q$ if the
following operator identity holds:
\begin{equation}
  \label{1.2}
  [Q,\ \partial^2_x-(\ve+V(x))]=R\,Q+P\,(\partial^2_x-(\ve+V(x)). 
\end{equation}

Here $[Q_1,\ Q_2]\equiv Q_1Q_2-Q_2Q_1$, $R$, $P$ are some first and
$n$-th order differential operators, correspondingly. The above
operator equality should be understood in the following way: the
differential operators in the left- and right-hand sides give the same 
result when acting on arbitrary $(n+2)$-times continuously
differentiable function $f(x)$.

Provided $R$ vanishes, condition (\ref{1.2}) is nothing else but a
criterion for equation (\ref{0.1}) to be invariant with respect to
the operator $Q$. In such a case, the operator $Q$ is a generalized
(high order) Lie symmetry operator. But given a condition $R\ne 0$,
the operator $Q$ corresponds to high order conditional symmetry of the
Schr\"odinger equation (\ref{0.1}).

It is easy to see that if an equation is conditionally invariant with
respect to the operator $Q$, then it is conditionally invariant with
respect to the operator $q(x)Q$ with an arbitrary sufficiently smooth
function $q(x)$. Consequently, without loss of generality we can
suppose that in (\ref{1.1}) $q_n(x)=1$ and consider differential
operators of the form
\begin{equation}
  \label{1.3}
  Q=\partial^n_x+\sum\limits_{j=0}^{n-1}q_{j}(x)\partial^j_x.
\end{equation}

As the coefficients of the operator $Q$ do not depend on $\ve$,
equality (\ref{1.2}) is only possible if $R=r(x)$ and $P=0$ with
some sufficiently smooth function $r(x)$. Consequently, the condition 
(\ref{1.2}) is rewritten to become

\begin{equation}
  \label{1.4}
  [Q,\ \partial^2_x - V(x)]=r(x)Q. 
\end{equation}

We call the Schr\"odinger operator {\em reducible} \/if there exist
linearly-inde\-pen\-dent functions $u_1(x),\ldots,u_n(x)$ and constants
$\Lambda_{jk}$ such that the conditions (\ref{0.2}) are fulfilled. Let
us note that this terminology is justified both from the point of view
of the classical representation theory and of the symmetry analysis of
differential equations. Indeed, conditions (\ref{0.2}) mean that the
representation space of the operator $S$ contains an invariant
subspace and, consequently the representation is reducible. On the
other hand, conditions (\ref{0.2}) ensure the reduction of the
differential equation (\ref{0.1}) to a system of algebraic equations.
We will prove an assertion which shows that this is not a simple
coincidence but a fundamental fact having a natural symmetry
interpretation.

\begin{theo}
  The Schr\"odinger operator $S=\p_x^2-V(x)$ is reducible if and only
  if there exists an $n$-th order differential operator $Q$ of the
  form (\ref{1.1}) such that equation (\ref{0.1}) is conditionally
  invariant with respect to $Q$.
\end{theo}
{\bf Proof.}\ {\em The necessity}.\ Let the operator $S=\p_x^2-V(x)$
be reducible. Then, the conditions (\ref{0.2}) hold. As functions
$u_j(x)$ are linearly independent, they form a fundamental system of
solutions of some $n$-th order linear ordinary differential equation
\cite{kam}.

We recall that fundamental system of solutions of an ordinary
differential equation is a maximal set of its particular solutions
such that any smooth solution can be represented as a linear
combination of these. Provided the order of the ordinary differential
equation in question is equal to $n$, any $n$ linearly independent
solutions of it form a fundamental system. Furthermore, having a
fundamental system of solutions we can reconstruct the corresponding
ordinary differential equation within a multiplication by a function
$r(x)$. Consequently, if we fix the coefficient of the $n$-th order
derivative to be equal to 1, then this equation is unique.

Thus, there exists the $n$-th order differential equation
\begin{equation}
\label{1.5}
u^{(n)}(x)+\sum\limits_{j=0}^{n-1}\tilde q_j(x)u^{(j)}(x)=0,
\end{equation}
such that the functions $u_j(x)$ form a fundamental system of its
solutions. 

We will prove that equation (\ref{0.1}) is conditionally invariant
with respect to the operator
\[
\tilde Q=\partial^n_x+\sum\limits_{j=0}^{n-1}\tilde
q_{j}(x)\partial^j_x. 
\]

By force of relations (\ref{0.2}) the following equalities hold:
\begin{eqnarray*}
&&[\tilde Q,\ \partial^2_x-V(x)]u_j(x) = \tilde Q \{(\partial^2_x -
V(x))u_j(x)\}\\
&&\qquad - (\partial^2_x - V(x))\{\tilde Q u_j(x)\} =
\tilde Q\left\{\sum\limits_{k=1}^n\Lambda_{jk}u_k(x)\right\}=0
\end{eqnarray*}
for any $j=1,2,\ldots,n$.

Thus, the functions $u_j(x)$ satisfy an ordinary differential equation
\begin{equation}
\label{1.6}
[\tilde Q,\ \partial^2_x-(\ve+V(x))] u(x)=0,
\end{equation}
whose order is easily established to be equal to $n$. Consequently,
its fundamental system of solutions consists of $n$ functions. Hence,
we conclude that the functions $u_j(x)$ form a fundamental system of
solutions of (\ref{1.6}). As an ordinary differential equation is
determined by its fundamental system uniquely within a multiplication
by a function $r(x)$, the relation hold
\[
[\tilde Q,\ \partial^2_x-(\ve+V(x))]=r(x)\tilde Q,
\]
which is the same as what was to be proved.

{\em The sufficiency}.\ Let the Schr\"odinger equation (\ref{0.1}) be
conditionally invariant with respect to the operator (\ref{1.3}),
which means that the condition (\ref{1.4}) is fulfilled. Consider an
equation
\begin{equation}
  \label{1.7}
  Qu(x)\equiv \left(\partial^n_x + \sum\limits_{j=0}^{n-1} q_{j}(x)
  \partial^j_x \right) u(x)=0
\end{equation}
as an ordinary differential equation for a function $u(x)$.
Clearly, its general solution is represented in the form
\begin{equation}
  \label{1.8}
  u(x)=\sum\limits_{j=1}^n\, C_j u_j(x),
\end{equation}
where $C_j$ are arbitrary constants and $u_1(x),\ldots,u_n(x)$ is a
fundamental system of solutions of $n$-th order ordinary differential
equation (\ref{1.7}).

From the condition (\ref{1.4}) it follows that the Schr\"odinger
operator $S=\partial^2_x-V(x)$ is a symmetry operator for the equation
(\ref{1.7}), i.e.\ it transforms each solution of equation (\ref{1.7})
into another solution of the same equation. Consequently, for any
$j=1,2,\ldots,n$ the function $\tilde u_j(x) = (\partial^2_x - V(x))
u_j(x)$ satisfy (\ref{1.7}). But by definition the fundamental system
of solutions of n-th order ordinary differential equation forms a
maximal set of its linearly independent solutions, which means that
any solution can be represented as a linear combination of functions
$u_j(x)$. Thus, there exist such constants $\Lambda_{jk}$ that
functions $u_j(x)$ satisfy relations (\ref{0.2}), whence it follows
that the corresponding Schr\"odinger equation is reducible. The
theorem is proved.

Note that the proof of theorem is, in fact, independent of the
specific form of the Schr\"odinger operator $S=\p_x^2-V(x)$. It is
straightforward to generalize Theorem 1 to the case of an
arbitrary $N$-th order differential operator
\begin{equation}
\label{new}
\tilde S = \sum\limits_{j=0}^{N}f_{j}(x)\partial^j_x.
\end{equation}

We give the corresponding assertion without proof.

\begin{theo}
  The operator $\tilde S$ in (\ref{new}) is reducible if and only if
  there exists $n$-th order differential operator $Q$ of the form
  (\ref{1.1}) such that equation $\tilde S \psi(x) =0$ is
  conditionally invariant with respect to $Q$.
\end{theo}

To illustrate the above statement we consider two examples. 
\vspace{2mm}

\noindent
{\em Example 1.}\ Consider the harmonic oscillator Schr\"odinger
equation 
\begin{equation}
  \label{1.9}
   \psi_{xx}=(\ve+x^2)\psi.
\end{equation}

As a direct check shows the $n$-th order differential operator
\begin{equation}
  \label{1.10}
  Q=(\p_x-x)^n
\end{equation}
satisfies the following commutation relation:
\[
[Q,\ \partial^2_x-(\ve+x^2)]=2nQ
\]
(the easiest way to prove the above formula is to use the mathematical 
induction method). 

Consequently, equation (\ref{1.9}) is conditionally invariant with
respect to the operator $Q$ and we can apply Theorem 1. Integrating
equation $Q\psi(x)=0$ yields a basis of the invariant space ${\cal
  V}_n$ of the Schr\"odinger operator $\p_x^2-x^2$
\[
{\rm e}^{-\frac{x^2}{2}},\quad x{\rm e}^{-\frac{x^2}{2}}, \quad
x^2{\rm e}^{-\frac{x^2}{2}},\ldots, \quad x^{n-1}{\rm
  e}^{-\frac{x^2}{2}}.
\]

It is readily seen that the above functions satisfy relations
(\ref{0.2}) with $V(x)=x^2$. Calculating eigenvalues ($\lambda_j$) and
eigenvectors ($\vec a^j$) of the corresponding matrix
$\|\Lambda_{jk}\|$ we obtain exact solutions of the Schr\"odinger
equation (\ref{1.9}) with $\ve=\lambda_1, \ldots, \lambda_m$ in the
form (\ref{0.4}).  
\vspace{2mm}

\noindent 
{\em Example 2.}\ Let us generalize the previous example as follows.
We are looking for the Schr\"odin\-ger equations (\ref{0.1})
conditionally invariant with respect to the $n$-th order operator
which can be represented as a power of the first order differential
operator, i.e.
\begin{equation}
\label{1.11}
Q=(a(x)\p_x + b(x))^{n+1}.
\end{equation}

By an appropriate transformation of the dependent and independent
variables 
\[
z=F(x),\quad \varphi (z)={\rm e}^{-\int G(x)dx}\psi(x),
\]
we can transform the operator $Q$ as follows:
\[
\tilde Q= \p_z^{n+1}.
\]

After rewriting the initial Schr\"odinger equation in the new
variables $z,\ \varphi(z)$ we get
\[
  f(z)\varphi_{zz}+g(z)\varphi_z+(h(z)-\ve)\varphi=0,
\]
where 
\begin{eqnarray*}
&&f(z)=(F'(x))^2,\quad g(z)= F''(x)+2 F'(x) G(x),\\
&&h(z) = - V(x) + G'(x) + G^2(x).
\end{eqnarray*}

Commutation relations (\ref{1.4}) now read as
\begin{equation}
  \label{1.12}
  [\p_z^{n+1},\ f(z)\p^2_z + g(z)\p_z + h(z)-\ve] = r(z)\p_z^{n+1}.
\end{equation}

Computing the commutator in the left-hand side (which is a simple
exercise in differential calculus) and equating coefficients of the
linearly independent operators $\p_z^j$ we conclude that the equation
(\ref{1.12}) is consistent if and only if the functions $f,\ g,\ h$
are polynomials in $z$ of the following form:
\begin{eqnarray*}
    h(z)&=&
{A_0} - n\left( {B_2} + (n - 1){C_3} \right) z + 
  {C_4}\left( n - 1 \right) n{z^2},\\
     g(z)&=&
{B_0} + {B_1}z + {B_2}{z^2} + 
  2{C_4}\left( 1 - n \right) {z^3},\\
     f(z)&=&
{C_0} + {C_1}z + {C_2}{z^2} + {C_3}{z^3} + {C_4}{z^4},
\end{eqnarray*} 
where $A_0, B_0,B_1,\ldots,C_4$ are arbitrary constants.

Returning back to the initial variables $x,\ u(x)$ we get the necessary
and sufficient conditions for the Schr\"odinger equation (\ref{0.1})
to be conditionally invariant with respect to an operator belonging to 
the class (\ref{1.11})
\begin{eqnarray*}
 &&   -V + G' + G^2=
{A_0} - n\left( {B_2} + (n - 1){C_3} \right) F + 
  {C_4}\left( n - 1 \right) n{F^2},\\
     &&F'' + 2 F' G=
{B_0} + {B_1}F + {B_2}{F^2} + 
  2{C_4}\left( 1 - n \right) {F^3},\\
     &&(F')^2=
{C_0} + {C_1}F + {C_2}{F^2} + {C_3}{F^3} +
{C_4}{F^4},
\end{eqnarray*} 
whence we derive the form of the potential $V(x)$
\begin{eqnarray} 
  V(x)&=&{{v_0} + {v_1}{\omega} + {v_2}{\omega}^2 + {v_3}{\omega}^3 +
    {v_4}{\omega}^4\over 16({C_0} + {C_1}{\omega} +
    {C_2}{{\omega}^2} + {C_3}{{\omega}^3} + {C_4}{{\omega}^4})},
  \label{1.13}\\ &&\nonumber\\ 
{v_0}&=&
4{{{B_0}}^2} - 16{A_0}{C_0} + 8{B_1}{C_0} - 
  8{B_0}{C_1} + 3{{{C_1}}^2} - 8{C_0}{C_2},\nonumber\\
{v_1}&=&
8{B_0}{B_1} + 16{B_2}{C_0}(n+1) - 16{A_0}{C_1} - 
  16{B_0}{C_2} + 4{C_1}{C_2}\nonumber\\
  && + 
  16{B_2}{C_0}n + 8{C_0}{C_3}(2{n^2}-2n-3),\nonumber\\
{v_2}&=&
4{{{B_1}}^2} + 8{B_0}{B_2} + 8{B_2}{C_1}(2n+1) - 
  16{A_0}{C_2} - 8{B_1}{C_2} + 4{{{C_2}}^2} \nonumber\\
  && - 24{B_0}{C_3} + 2{C_1}{C_3}(8n^2-8n-3) -
  16{C_0}{C_4}n(n+2),\nonumber\\  
{v_3}&=&
8{B_1}{B_2} - 16{A_0}{C_3} - 16{B_1}{C_3} -
16{B_0}{C_4}(n+1) + 16{B_2}{C_2}n\nonumber\\ 
  &&  - 
  16{B_0}{C_4}n - 8{C_1}{C_4}(2n^2+2n-1) + 
  4{C_2}{C_3}(4n^2 -4n+1),\nonumber\\
{v_4}&=&
4{{{B_2}}^2} + 8{B_2}{C_3}(2n-1) - 
  16{A_0}{C_4} - 8{B_1}{C_4}(2n+1) + \nonumber\\
  && + {{{C_3}}^2}(16n^2-16n+3) + 
  8{C_2}{C_4}(1-2{n^2}) \nonumber
\end{eqnarray}
and of the function $G(x)$
\[
G(x)={{2{B_0} - {C_1} + 2({B_1} - {C_2}){\omega} + 
     (2{B_2} - 3{C_3}){{\omega}^2} -
     4{C_4}n{{\omega}^3}}\over {4{\sqrt{{C_0} +
         {C_1}{\omega} + {C_2}{{\omega}^2} + {C_3}{{\omega}^3} +
         {C_4}{{\omega}^4}}}}}.
\]

In the above formulae $\omega(x)$ is an elliptic function determined
by the quadrature
\[
\int\limits^{\omega(x)}{d\tau \over \sqrt{{C_0} + {C_1}\,{\tau} +
    {C_2}\,{{\tau}^2} + {C_3}\,{{\tau}^3} + {C_4}\,{{\tau}^4}}}=x.
\]

Furthermore, exact solutions of the Schr\"odinger equation with the
potential (\ref{1.13}) read as
\[
\psi (x)={\rm e}^{\int G(x)dx}\sum\limits_{j=0}^{n}a_j\omega(x)^j,
\]
where $\vec a =(a_0,a_1,\ldots, a_n)$ is an eigenvector of some
$(n+1\times n+1)$ constant matrix whose entries are linear
combinations of the parameters $A_0$,$B_0$,$B_1$,$\ldots$, $C_4$ (we
omit the corresponding formulae).

Thus, we arrived at the nine-parameter family of quasi exactly
solvable Schr\"odinger equations obtained by {\sc Turbiner} and {\sc
  Shifman} within the framework of their Lie algebraic approach
\cite{tur,shi} and by {\sc Ushveridze} by means of a more general
analytic approach. A detailed account of properties of the
Schr\"odinger equation with potentials (\ref{1.13}) can be found in
the monograph \cite{ush}. We restrict ourselves to noting that if we
choose in the above formulae $B_2=C_3=C_4=0$, then the potential
$V(x)$ does not depend on $n$ (the order of the operator $Q$) and,
consequently, the corresponding Schr\"odinger equation is exactly
solvable. Thus, the well-known six-parameter family of exactly solvable
Schr\"odinger equations is obtained. In particular, choosing
$C_1=C_2=C_3=C_4=0,\ C_0=1$ yields the harmonic oscillator
Schr\"odinger equation (\ref{1.9}).

Now let us pass from particular examples to the general case in order
to examine which constraints are imposed on the coefficients of the
operator (\ref{1.3}) by the requirement of conditional invariance
(\ref{1.4}). In order to compute the commutator on the left-hand side
of (\ref{1.4}) we use the following identity:
\begin{equation}
\label{1.14}
[\p_x^k,\, f(x)]=\sum\limits_{j=0}^{k-1}{\rm C}_k^{j}f^{(k-j)}(x)
\p_x^j,\quad k\in {\bf N},
\end{equation}
where ${\rm C}_j^k=k!(j!(k-j)!)^{-1}$ are binomial coefficients, which
is established by the mathematical induction method with the help of
the evident identity
\[
[\p_x^{k+1},\, f(x)]\equiv \p_x[\p_x^{k},\, f(x)] + f'(x)
\p_x^{k}. 
\]

Taking into account formula (\ref{1.14}) we rewrite relation
(\ref{1.4}) as follows:
\begin{eqnarray*}
&&- \sum\limits_{j=0}^{n-1}\,{\rm C}^{j}_{n}\, V^{(n-j)}\,
\p_x^j-\sum\limits_{i=1}^{n-1} 
\sum\limits_{j=0}^{i-1}\, q_i\, {\rm C}^{j}_{i}\, V^{(i-j)}\,
\p_x^j -\sum\limits_{j=0}^{n-1}\, (2q'_j\p_x + q''_j)\,\p_x^j\\ 
&&\phantom{- \sum\limits_{j=0}^{n-1}}= r(x) \left (\p_x^n +
\sum_{j=0}^{n-1}\,q_j\p_x^j\right ).  
\end{eqnarray*}

Comparing the coefficients of $\p_x^n$ on the left- and right-hand
sides of the above equation we conclude that $r(x)=-2q'_{n-1}$.
Comparing the coefficients of the linearly independent operators
$\p_x$,\ $\p_x^2$,$\ldots$, $\p_x^{n-1}$ we arrive at the following
system of nonlinear ordinary differential equations for the functions
$q_0(x)$,\ $q_1(x)$,$\ldots$, $q_{n-1}(x)$,\ $V(x)$:
\begin{equation}
\label{1.15}
2q'_{j-1} + q''_j + {\rm C}^{j}_{n}\, V^{(n-j)} +
\sum\limits_{i=j+1}^{n-1} \,q_i\, {\rm C}^{j}_{i}\, V^{(i-j)}
-2q'_{n-1}\,q_j = 0,
\end{equation}
where $j=0,1,\ldots, n-1$ and by convention $q_{-1}\stackrel{\rm
  def}{=}0,\ q_{n}\stackrel{\rm def}{=}1$.

Thus, we have $n$ equations for $n+1$ functions, which means that the
system (\ref{1.15}) is under-determined. As an immediate consequence
of this fact we conclude that {\em any} \/Schr\"odinger equation
(\ref{0.1}) is reducible. Indeed, fixing in an arbitrary way a
function $V=V(x)$ yields a second-order system of $n$ ordinary
differential equations for $n$ functions $q_0(x)$,\ $q_1(x)$,$\ldots$,
$q_{n-1}(x)$. Each solution of such a system gives rise to an operator
$Q$ satisfying by construction condition (\ref{1.4}). Consequently,
the conditions of Theorem 1 can be fulfilled with any choice of
the potential $V(x)$.

Let us demonstrate how the results obtained can be used to study the
spectral properties of the Schr\"odinger equation. Remarkably, to this
end we do not need an explicit form of solution of the system of
nonlinear ordinary differential equations (\ref{1.15}). It suffices to
know initial values of the functions $q_j(x)$ and of their first
derivatives $q'_j(x)$ at some point $x=x_0$ $\in$ ${\bf R}$. We denote
these as follows
\begin{equation}
\label{1.16}
q_j(x_0)=A_{j+1},\quad q'_{j}(x_0)=B_{j+1},\quad j=0,1,\ldots,n-1.
\end{equation}

All the information about spectral properties of the Schr\"odinger
operator restricted to an invariant space ${\cal V}_n$ with basis
elements $u_1(x)$\ ,$u_2(x)$,\ $\ldots$,$u_n(x)$ is contained in the
matrix $\|\Lambda_{jk}\|$, which determines a transformation law
(\ref{0.2}) for the functions $u_j(x)$ with respect to the action of
the Schr\"odinger operator $S$.

Let $Q$ be a differential operator of the order $n$ satisfying
condition (\ref{1.4}) with some choice of the function $V(x)$. Then,
by force of Theorem 1, an invariant space ${\cal V}_n$ of the
corresponding Schr\"odinger operator $S$ is spanned by the fundamental
system of solutions of the $n$-th order ordinary differential equation
\begin{equation}
\label{1.17}
u^{(n)}(x)+\sum\limits_{j=0}^{n-1} q_j(x)u^{(j)}(x)=0.
\end{equation}

We denote this system as $\{u_1(x)$,\ $u_2(x)$,\
$\ldots$,$u_n(x)\}$. As any $n$ linearly independent solutions of
equation (\ref{1.17}) form a fundamental system of solutions, there is
a freedom in the choice of the functions $u_j(x)$. We fix these by
imposing initial conditions. 

Let $\|L_{jk}\|$ be a constant non-singular $n\times n$
matrix. Consider the following $n$ Cauchy problems
\begin{eqnarray}
&&u_k^{(n)}(x)+\sum\limits_{j=0}^{n-1} q_j(x)u_k^{(j)}(x)=0,\quad
k=1,\ldots, n \label{1.18}\\
&& u_k^{(j-1)}(x_0)=L_{kj},\quad k,j=1,\ldots,n.\nonumber
\end{eqnarray}
 
It is well-known from the general theory of linear differential
equations that the above system has a unique solution, and what is
more, this solution yields a fundamental system of solutions of
equation (\ref{1.17}).

Differentiating relations (\ref{0.2}) $n-1$ times with respect to $x$
and excluding the $n$-th and the $(n+1)$-th derivatives of the
functions $u_j(x)$ with the help of equations (\ref{1.18}) we
arrive at the following relations:
\begin{eqnarray}
\sum\limits_{k=1}^{n}\Lambda_{jk}u_k^{(i)}&=&u_j^{(i+2)} -
\sum\limits_{k=0}^{i} {\rm C}_{i}^{k} V^{(i-k)}u_j^{(k)},\nonumber\\ 
\sum\limits_{k=1}^{n}\Lambda_{jk}u_k^{(n-2)} &=&
-\sum\limits_{k=0}^{n-1} q_ku_j^{(k)} - \sum\limits_{k=0}^{n-2} {\rm  
  C}_{n-2}^{k} V^{(n-k-2)}u_j^{(k)},\label{1.19}\\
\sum\limits_{k=1}^{n}\Lambda_{jk}u_k^{(n-1)}&=& 
\sum\limits_{k=0}^{n-1}\left (q_{n-1}q_k - q'_k - q_{k-1} - {\rm
  C}_{n-1}^{k} V^{(n-k-1)}\right )u_j^{(k)}, \nonumber 
\end{eqnarray}
where $j=1,\ldots,n$,\ $i=0,\ldots,n-3$ and as above
$q_{-1}\stackrel{\rm def}{=}0$.

Choosing $x=x_0$ in (\ref{1.19}) yields
\[
\sum\limits_{k=1}^{n}\Lambda_{jk}L_{ki} = \sum\limits_{k=1}^{n}
L_{jk}R_{ki},\quad j,i=1,\ldots,n,
\]
where
\begin{eqnarray}
R_{k\,i}&=&\delta_{k\ {i+2}} -
{\rm C}_{i-1}^{k-1}V^{(i-k)}(x_0),\quad i=1,\ldots,n-2, \nonumber\\
R_{k\, n-1}&=& - A_{k} - {\rm C}_{n-2}^{k-1}V^{(n-k-1)}(x_0),
\label{1.20}\\ 
R_{k\, n}&=& A_nA_k - B_k -A_{k-1} -
{\rm C}_{n-1}^{k-1}V^{(n-k)}(x_0). \nonumber
\end{eqnarray}
 
In formulae (\ref{1.20}) the index $k$ runs from $1$ to $n$, $A_k$,\ 
$B_k$ are constants defined by (\ref{1.16}) and $\delta_{k\, j}$ is
the Kronecker symbol.

Rewriting the formulae obtained in the matrix form we have $\Lambda\,
L= L\, R$, where $\Lambda,\ L,\ R$ are constant $n\times n$ matrices
with entries $\Lambda_{jk},\ L_{jk},\ R_{jk}$, respectively. Hence, we
derive the explicit form of the matrix $\Lambda$
\begin{equation}
  \label{1.21}
  \Lambda = L\, R\, L^{-1}.
\end{equation}

Thus we have proved the following assertion.

\begin{theo}
  Let $L$ be an arbitrary invertible $n\times n$ matrix, and
  $A_1,\ldots,A_n$, $B_1,\ldots,B_n$ be arbitrary constants. Then, for
  any choice of the function $V(x)$ there exist $n$ functions
  $u_1(x)$,$\ldots$, $u_n(x)$ such that the relations (\ref{0.2}) hold,
  $\Lambda_{jk}$ being the entries of the $n\times n$ matrix given by
  formulae (\ref{1.20}), (\ref{1.21}).
\end{theo}

The above theorem has as a consequence the following important
assertion which describes a finite part of the spectrum of the
Schr\"odinger operator $S$.

\begin{theo}
  Let $\lambda_1,\ldots,\lambda_m$ be distinct eigenvalues of the
  matrix $\|R_{jk}\|$ determined by formulae (\ref{1.20}). Then,
  for any choice of parameters $A_j,\ B_j$ there exist
  linearly independent functions $\psi_1(x),\ldots,\psi_m(x)$
  satisfying the Schr\"o\-din\-ger equation (\ref{0.1}) with
  $\ve=\lambda_1$,\ $\ve=\lambda_2$, $\ldots$, $\ve=\lambda_m$,
  correspondingly.
\end{theo}

The proof follows from Theorem 2 if one takes into account that the
matrix $\Lambda$ is similar to $R$ and, consequently, has the same
eigenvalues $\lambda_1$,\ $\lambda_2$, $\ldots$, $\lambda_m$. The
explicit form of the functions $\psi_1(x),\ldots,\psi_m(x)$ is given
by the formula (\ref{0.4}), where $\vec a^{j}=(a_1^{j}, a_2^{j},
\ldots, a_n^{j}),\ j=1,\ldots,m$ are eigenvectors of the matrix
$\Lambda$ corresponding to the eigenvalues $\lambda_1$,\ $\lambda_2$,
$\ldots$, $\lambda_m$. 

Thus, using the conditional symmetry approach we were able not only to
calculate the spectrum of the Schr\"odinger operator $S=\p_x^2-V(x)$
but also to construct in an explicit form a $2n$-parameter family of
matrix representations of $S$.

As is seen from formulae (\ref{1.20}), (\ref{1.21}), there is
a large freedom in choice of the matrix $\Lambda$. First, it depends
on $2n$ arbitrary constants $A_j,\ B_j$, which fix a fundamental
system of solutions of ordinary differential equation (\ref{1.17}).
Consequently, choosing specific constants $A_j,\ B_j$ means fixing a
representation space ${\cal V}_n$. Secondly, it contains an arbitrary
constant $n\times n$ matrix $L$. The appearance of the matrix $L$ in the
definition of $\Lambda$ reflects a freedom in choosing a basis of the
representation space ${\cal V}_n$. Indeed, if $u_1(x),\ldots, u_n(x)$
is a basis of the space ${\cal V}_n$, then the functions
$\sum^n_{k=1}L_{jk}u_k(x)$ also form a basis with an arbitrary
invertible constant $n\times n$ matrix $\|L_{jk}\|$. That is why,
choosing a specific matrix $L$ results in fixing a basis of the
representation space $u_1(x)$, $\ldots$, $u_n(x)$. This freedom can be
used, in particular, to obtain an orthogonal basis for ${\cal V}_n$
(to this end one should apply the standard Gram-Schmidt
orthogonalization procedure).

But representations of the Schr\"odinger operator in these basises are
equivalent, which is readily seen from the formula (\ref{1.21}).

Thus, it is established that {\em any} \/Schr\"odinger equation has
$n$-dimensional invariant spaces ${\cal V}_n$ with arbitrary $n \in
{\bf N}$. Furthermore, we have constructed the $2n$-parameter family
of matrix representations of the corresponding Schr\"odinger operator
in these spaces (the formulae (\ref{1.20}), (\ref{1.21})). But to
obtain an explicit form of the basis of ${\cal V}_n$ we still have
\begin{itemize}
\item{to integrate system of nonlinear ordinary differential equations
    (\ref{1.15}), and}
\item{to construct the general solution of the n-th order ordinary
    differential equation (\ref{1.17}).}
\end{itemize}

We will demonstrate that using a simple trick we may avoid the
necessity to integrate equation (\ref{1.17}). The said trick is based
on the fact that we need not all solutions of (\ref{1.17}) but only
those which simultaneously satisfy the initial Schr\"odinger equation
(\ref{0.1}). This means that we have to solve the following
over-determined system of two ordinary differential equations:
\begin{equation}
\left\{\begin{array}{l}
  \label{1.22}
  \psi_{xx}=(\ve+V(x))\psi,\\[2mm]
  \psi^{(n)}(x)+\sum\limits_{j=0}^{n-1} q_j(x)\psi^{(j)}(x)=0.
\end{array}\right.
\end{equation}

Using the first equation and its differential consequences up to the
order $n-2$ we can exclude from the second equation all the
derivatives of the function $\psi$ of the order $j>1$ and rewrite
system (\ref{1.22}) in the following equivalent form:
\begin{equation}
\left\{ \begin{array}{l}
  \label{1.23}
  \psi_{xx}=(\ve+V(x))\psi,\\[2mm]
\left (\sum\limits_{i=0}^N\, a_i(x)\ve^i\p_x + \sum\limits_{i=0}^N\,
  b_i(x)\ve^i \right)\psi=0,
\end{array}\right.
\end{equation}
where $N=[\frac{n}{2}]$,\ $a_i(x),\ b_i(x)$ are linear combinations of
the functions $q_j(x)$ with coefficients depending on $V(x)$ and its
derivatives (and, consequently, independent of $\ve$) and,
furthermore, $a_N=1$ if $n=2N+1$ and $a_N=0,\ b_N=1$ if $n=2N$.

The compatibility condition for the above system reads as

\begin{equation}
\label{1.24}
\left( {\sum_{i=0}^N\, b_i(x)\ve^i\over \sum_{i=0}^N\,
    a_i(x)\ve^i}\right )_{x}- \left( {\sum_{i=0}^N\,
    b_i(x)\ve^i\over \sum_{i=0}^N\, a_i(x)\ve^i}\right )^2 + V(x) +\ve
  = 0.
\end{equation}

As functions $a_i,\ b_i$ are independent of $\ve$, coefficients of the 
powers of $\ve$ should be independent of $x$. This requirement yields
\begin{enumerate}
\item[{1)}]{$2N+1$ ordinary differential equations
\begin{equation}
\label{1.25}
\sum\limits_{i+j=k}\, \left( b'_ia_j - b_ia'_j - b_ib_j + V a_i
a_j\right ) 
+\sum\limits_{i+j=k-1}\,a_ia_j = C_k,
\end{equation}
where $k=0,1,\ldots,2N$, for $2N+2$ functions $a_0,\ldots,a_{N-1}$,\ 
$b_0,\ldots, b_N$,\ $V$, provided $n=2N+1$, or}
\item[{2)}]{$2N$ ordinary differential equations of the form (\ref{1.25})
    with $k$ taking the values $0,1$, $\ldots$, $2N-1$ for $2N+1$ functions
    $a_0,\ldots,a_{N-1}$,\ $b_0,\ldots, b_{N-1}$,\ $V$, provided
    $n=2N$.}
\end{enumerate}

In (\ref{1.25}) $C_j$ are arbitrary constants.

Provided conditions (\ref{1.25}) are fulfilled, the compatibility
condition (\ref{1.24}) reduces to an algebraic equation
\begin{enumerate}
\item[{1)}]{under $n=2N+1$
\begin{equation}
\label{1.26}
\ve^{2N+1}+\sum\limits_{j=0}^{2N}\, C_j \ve ^k = 0,
\end{equation}}
\item[{2)}]{under $n=2N$
\begin{equation}
\label{1.27}
\ve^{2N}+\sum\limits_{j=0}^{2N-1}\, C_j \ve ^k = 0
\end{equation}}
\end{enumerate}
and the general solution of system (\ref{1.23}) is given by the
quadrature 
\begin{equation}
\label{1.28}
\psi(x)=\exp \left \{-\int \left ({\sum_{i=0}^N\, b_i(x)\ve^i\over 
  \sum_{i=0}^N\, a_i(x)\ve^i}\right )\, dx\right \}.
\end{equation}

It should be noted that equations (\ref{1.26}), (\ref{1.27}) are
nothing else but characteristic equations for the matrix $R$ defined
by (\ref{1.20}). Their solutions $\lambda_1,\ldots,\lambda_m$ are
eigenvalues and $\psi_1(x)=\psi(x)|_{\ve=\lambda_1}$, $\ldots$,
$\psi_m(x)=\psi(x)|_{\ve=\lambda_m}$ are eigenfunctions of the
corresponding Schr\"odinger operator $S=\p_x^2-V(x)$.

\section*{III.\ Symmetry and exact solvability}

With the best of our knowledge the first paper, where a systematic
study of high order Lie symmetries of the Schr\"odinger equations with
non-vanishing potentials has been undertaken is the one by {\sc
  Nikitin, Onufriychuk} and {\sc Fushchych} \cite{nik1}. In
particular, for several one-dimensional exactly solvable models
third-order symmetry operators were constructed. Furthermore, it was
conjectured that exact integrability of the Schr\"odinger equation
(\ref{0.1}) is intimately connected with its symmetry properties. This
conjecture has been confirmed in \cite{nik2}, where a number of
exactly solvable potentials were obtained by means of third-order
symmetries of (\ref{0.1}) and, furthermore, a method for integrating
the corresponding Schr\"odinger equations was suggested. Using a
technique developed in the previous section we will demonstrate how to
derive exact integrability of equation (\ref{0.1}) from its symmetry
properties in the class of arbitrary order symmetry operators.

The $n$-th order operator $Q$ of the form (\ref{1.1}) is a symmetry
operator of equation (\ref{0.1}) if the condition (\ref{0.2}) is
satisfied with $r(x)=0$. As coefficients of $Q$ are independent of
$\ve$, equality (\ref{0.2}) is only possible when $P\equiv 0$.
Computing the commutator on the left-hand side of (\ref{0.2}) with
$r=0,\ P=0$ and equating to zero the coefficients of $\p_x^{n+1}$ and
$\p_x^{n}$ we conclude that $q_n(x)={\rm const},\ q_{n-1}(x)={\rm
  const}$. Consequently, any $n$-th order symmetry operator admitted
by the Schr\"odinger equation (\ref{0.1}) can be represented in the
form (\ref{1.1}) with $q_n=C={\rm const}\ne 0,\ q_{n-1}=C_0={\rm
  const}$.

First, following \cite{nik2} we will consider in more detail the
third-order symmetry operators admitted by the Schr\"o\-din\-ger
equation (\ref{0.1}). These are the lowest order symmetry operators
not equivalent to usual first-order Lie symmetries. The general form
of a third-order symmetry operator is as follows (we choose $C=1$)
\begin{equation}
  \label{2.1}
  Q=\p_x^3+C_0\p_x^2+q_1(x)\p_x+q_0(x),
\end{equation}
where $q_0(x),\ q_1(x)$ are sufficiently smooth functions to be
determined from the invariance condition (\ref{0.2}) with $n=3,\ 
r=P=0$. A short computation yields the following expressions for the
coefficients of the operator $Q$
\[
  q_0(x)={C_2} - {C_0}V(x) - {{3}\over 4}V'(x),\quad
  q_1(x)={C_1} - {{3}\over 2}V(x),
\]
where $V(x)$ is an arbitrary solution of the third-order nonlinear
ordinary differential equation
\begin{equation}
\label{2.2}
 -4\,{C_1}\,V' + 6\,V\,V' - V^{(3)}=0. 
\end{equation}

Integrating twice the above equation we arrive at the first-order
ordinary differential equation integrable in elliptic functions
\begin{equation}
\label{2.3}
{C_4} + 2\,{C_3}\,V(x) + 4\,{C_1}\,{{V(x)}^2} - 2\,{{V(x)}^3} +
{{V'(x)}^2}=0.
\end{equation}

It was established in \cite{nik2} that particular cases of almost all
exactly solvable potentials which can be expressed in elementary
functions, such as the trigonometric and hyperbolic P\"oschel-Teller,
Eckart, Kratzer potentials, potential well of finite and infinite
depth, are obtained as solutions of the equation (\ref{2.3}).

This fact imply an existence of an intimate connection between high
order Lie symmetries and exact solvability of the Schr\"odinger
equations. In what follows we will show that this is not simply a
conjecture but a fundamental fact making it possible to classify
exactly solvable models and to construct their exact solutions in an
explicit form (see, also \cite{nik2,nik3}).

It is straightforward to check that if $Q$ is a Lie symmetry of the
Schr\"o\-din\-ger equation (\ref{0.1}), then $\tilde
Q=Q+P\,(\p_x^2-V(x)-\ve)$, where $P$ is an arbitrary differential
operator, is also its Lie symmetry. Furthermore, systems
of ordinary differential equations 
\[
(\p_x^2-V(x)-\ve)\psi(x)=0,\quad Q\psi(x)=0
\]
and 
\[
(\p_x^2-V(x)-\ve)\psi(x)=0,\quad \tilde Q\psi(x)=0
\]
are equivalent in a sense that they have the same solutions.
Making use of these facts we can reduce the $n$-th order symmetry
operator $Q$ to a first-order symmetry operator, the coefficients of
which are $N$-th order polynomials in $\ve$ (we will preserve the same
designation $Q$ for the reduced operator)
\begin{equation}
  \label{2.4}
  Q=a(x,\ve)\p_x+b(x,\ve)\equiv
  \left ( \sum\limits_{j=0}^N\,a_j(x)\ve^j \right)\p_x +
    \sum\limits_{j=0}^Nb_j(x)\ve^j.
\end{equation}

From the invariance condition for the reduced operator
\[
[\p_x^2-V(x)-\ve,\ a(x,\ve)\p_x+b(x,\ve)]=R(x,\ve)(\p_x^2-V(x)-\ve)
\]
we get a system of determining equations for the coefficients of $Q$
\begin{eqnarray*}
&&a''(x,\ve) +2 b'(x,\ve)=0,\\
&& b''(x,\ve)+a(x,\ve) V'(x)+2 a'(x,\ve)(V(x)+\ve) = 0,
\end{eqnarray*}
where primes denote differentiation with respect to $x$.

Splitting the above equations by powers of $\ve$ with subsequent
integrating yields
\begin{equation}
\label{2.5}
\begin{array}{l}
b_j(x)=-{\ds \frac{1}{2}} a'_j(x) + B_j,\quad a_N(x)=A_N, \\[2mm]
a_{j-1}(x)={\ds \frac{1}{4}} a''_j(x)
-V(x)a_j(x) + {\ds \frac{1}{2}}{\ds \int} V'(x) a_j(x)\,dx + A_{j-1}.
\end{array}
\end{equation}

In (\ref{2.5}) $A_{-1},\ A_j,\ B_j$ are arbitrary constants,\ 
$j=0,1,\ldots,N$,\ $a_{-1}(x)\stackrel{\rm def}{=}0$.

Thus, the problem of describing $n$-th order symmetry operators of the
Schr\"odinger equation is reduced to solving the recurrent relations
\begin{equation}
\label{2.6}
a_{j-1}(x)={\ds \frac{1}{4}} a''_j(x)
-V(x)a_j(x) + {\ds \frac{1}{2}}{\ds \int} V'(x) a_j(x)\,dx + A_{j-1}
\end{equation}
with $a_N(x)=A_N={\rm const}$,\ $a_{-1}(x)\stackrel{\rm def}{=}0$,\ 
$j=N,N-1,\ldots,0$.

The first $N$ relations ($j=N,N-1,\ldots,1$) are solved by subsequent
integrations yielding the expressions for the functions
$a_0(x),\ldots,a_{N-1}(x)$ via the function $V(x)$ and its
derivatives. Substituting these results into the last equation ($j=0$)
we arrive at the $2N$-th order nonlinear ordinary differential
equation for the function $V(x)$. It will be shown that any solution
of this equation gives rise to an exactly solvable Schr\"odinger
equation (\ref{0.1}).

To reveal the structure of the equation in question we introduce the
new functions $U_0(x)$,\ $U_1(x)$,$\ldots$ by the following recurrence
relation:
\begin{equation}
  \label{2.7}
  U_{j}(x)=X\, U_{j-1}(x)\equiv \left (\frac{1}{4}\p_x^2 - V(x) +
  \frac{1}{2}\p_x^{-1}V'(x)\right)\, U_{j-1},\quad j=0,1,\ldots
\end{equation}
where $\p_x^{-1}$ denotes integration with respect to $x$ and
$U_{-1}(x) \stackrel{\rm def}{=}1$.

Formally, a definition of functions $U_j(x)$ contains a quadrature but
integrating relations (\ref{2.7}) successively we can get rid of it
for any $j=1,2,3,\ldots$. Below, we adduce expressions of the
functions $U_j(x)$ for $j=0,1,2,3$.
\begin{eqnarray*} 
 U_0(x)&=&
-{1\over 2}{V(x)},\\
 U_1(x)&=&
{1\over 2^3}\left({3\,{{V(x)}^2} - V''(x)}\right),\\
 U_2(x)&=&
{1\over {2^5}}\left({-10\,{{V(x)}^3} + 5\,{{V'(x)}^2} +
    10\,V(x)\,V''(x) - V^{(4)}(x)}\right),\\
 U_3(x)&=&
  {1\over {2^7}}\left(35\,{{V(x)}^4} - 70\,V(x)\,{{V'(x)}^2} -
  70\,{{V(x)}^2}\,V''(x) + 21\,{{V''(x)}^2}\right.\\
  &&\phantom{{1\over {2^7}} )}\left. + 28\,V'(x)\,V^{(3)}(x) +
  14\,V(x)\,V^{(4)}(x) - V^{(6)}(x)\right).
\end{eqnarray*}

Now we can solve the first $N$ relations of (\ref{2.6}) in terms of
the functions $U_j(x)$
\begin{equation}
\label{2.8}
a_{N-j}(x)=\sum\limits_{k=0}^{j-1}\,A_{N-k}U_{j-k-1}(x) +
A_{N-j},\quad j=1,\ldots,N.
\end{equation}

Inserting the above expressions into the last equation of
(\ref{2.6}) we get
\begin{equation}
\label{2.9}
A_{-1} + \sum\limits_{k=0}^{N}\,A_{N-k}U_{N-k}(x)=0
\end{equation}
(when deriving the above equation we take into account that by
convention $a_{-1}(x)=0$).

Equation (\ref{2.9}) is the necessary and sufficient condition for the
Schr\"odinger equation (\ref{0.1}) to be invariant with respect to the
first-order operator (\ref{2.4}), whose coefficients are polynomials
in $\ve$ of the order $N$. But it is easy to see that given a
condition (\ref{2.9}), equation (\ref{0.1}) admits operator of the
form (\ref{2.4}), whose coefficients are polynomials in $\ve$ of an
arbitrary order $N'>N$.

Indeed, the invariance conditions for the operator
\[
  Q= \left ( \sum\limits_{j=0}^{N'}\,a_j(x)\ve^j \right)\p_x +
    \sum\limits_{j=0}^{N'}b_j(x)\ve^j
\]
have the form (\ref{2.5}) with $N=N'$. Coefficients $a_j(x)$ with
$j=N',N'-1,\ldots,N'-N$ are given by formulae (\ref{2.8}), where
one should replace $N$ by $N'$, and the remaining coefficients by
force of relation (\ref{2.9}) read as
\[
a_j(x)=\sum\limits_{k=0}^{N-1}\,\tilde A_{jk}U_k(x) + A_j,\quad
j=0,1,\ldots, N'-N-1,
\]
where $A_0,\ldots,A_{N'-N-1}$ are arbitrary constants,\ $\tilde
A_{jk}$ are constants expressed via $A_{N'-N}$, $\ldots$, $A_{N'}$.

Substituting these results into the last ($j=0$) equation from
(\ref{2.8}) yields
\[
A_{-1} + A_0 U_0(x) + \sum\limits_{k=0}^{N-1}\,\tilde A_{0k}U_{k+1}(x)
= 0,
\]
whence we conclude that, provided $A_{-1}=A_0=0$,\ $\tilde A_{0k}=0$,
$k=0,1,\ldots,N-1$, the invariance conditions are satisfied.

Thus, if equation (\ref{2.9}) is fulfilled with some $N\in {\bf N}$,
then the corresponding Schr\"odinger equation admits arbitrary order
Lie symmetries and, consequently, is exactly solvable.

A general solution of the Schr\"odinger equation invariant under the
operator $Q$ is given by (\ref{1.28}). Substituting formula
(\ref{1.28}) into equation (\ref{0.1}), where $V(x)$ is an
arbitrary solution of (\ref{2.9}), results in a $(2N+1)$-th order
algebraic equation for $\ve$. Its solutions $\ve=\lambda_1$, $\ldots$,
$\ve=\lambda_m$ are eigenvalues of the Schr\"odinger operator.
Corresponding eigenfunctions are obtained if we insert
$\ve=\lambda_1$, $\ldots$, $\ve=\lambda_m$ into (\ref{1.28}).

Summing up, we conclude that any solution $V(x)$ of (\ref{2.9}) gives
rise to an exactly solvable Schr\"odinger equation. In what follows
it will be established that the more strong assertion holds. Namely,
if $V(x)$ is a solution of (\ref{2.9}) with some $N$, then the
corresponding Schr\"odinger equation may have an {\em arbitrary}
spectrum.
\begin{theo}
Let $V=V(x)$ be a solution of ordinary differential equation
(\ref{2.9}) with some fixed $N\in {\bf N}$. Then, the Schr\"odinger
equation (\ref{0.1}) is exactly solvable and, moreover, the
Schr\"odinger operator $S=\p_x^2-V(x)$ may have an arbitrarily
prescribed spectrum.
\end{theo}
{\bf Proof.}\ We will give the principal steps of the proof omitting
technical details.  As the potential $V(x)$ satisfies equation
(\ref{2.9}), the corresponding Schr\"odinger equation admits a Lie
symmetry of the form (\ref{2.4}). Excluding from (\ref{2.4}) the
parameter $\ve$ we recover symmetry operator $Q$ of the order
$n=2N+1$ which commutes with the Schr\"odinger operator $S$.
Next, we construct an operator $Q_1=Q+f(\ve)$, where $f(\ve)$ is an
arbitrary smooth function.  Evidently, $Q_1$ commutes with $S$ and,
consequently, is a symmetry operator of the Schr\"odinger equation.
This means that conditions of Theorem 1 are fulfilled and a
fundamental system of solutions of ordinary differential equation
\[
Q_1u(x)=0
\]
forms a basis of the invariant space ${\cal V}_n$ of the corresponding
Schr\"odinger operator $S$. Representation of $S$ in the space ${\cal
  V}_n$ is given by the formulae (\ref{1.20}), (\ref{1.21}), where
$A_1=a+f(\ve)$ and parameters $a,\ A_2$, $\ldots$, $A_n$,\ $B_1$,
$\ldots$, $B_N$ are independent of $f(\ve)$. Eigenvalues of the
operator $S$ are solutions of the characteristic equation for the
matrix $R$ having the entries (\ref{1.20}), i.e.\ of the equation
\[
{\rm det}\,\|R_{jk}-\ve \delta_{jk}\|=0.
\]

It is not difficult to become convinced of that the above equation can
be represented in the form
\[
p_0(\ve)+p_1(\ve)f(\ve)+p_2(\ve)f(\ve)^2=0,
\]
where $p_0,\ p_1,\ p_2$ are polynomials in $\ve$ of the order not
higher than $n$.

Consequently, zeros of the function 
\[
F(\ve)=p_0(\ve)+p_1(\ve)f(\ve)+p_2(\ve)f(\ve)^2
\]
are eigenvalues of the Schr\"odinger operator.

As $f(\ve)$ is arbitrary, the function $F(\ve)$ may have an
arbitrarily prescribed set of zeros and, thus, a spectrum of the
initial Schr\"odinger equation (\ref{0.1}) may be arbitrary.

As an illustration, we will consider the simplest case when there
exists such $N_1$ that $U_{N_1}(x)=0$. In such a case, the
coefficients of the symmetry operator (\ref{2.4}) with $N=kN_1,\ k\in
{\bf N}$ are easily shown to be
\begin{eqnarray*}
a(x,\ve)&=&\left(\sum\limits_{j=-1}^{{N_1}-1}\,U_j(x)\ve^{{N_1}-j-1}
\right) \left(\sum\limits_{j=0}^k\,A_j\ve^j\right),\\
b(x,\ve)&=&-\frac{1}{2}\left(\sum\limits_{j=-1}^{{N_1}-1}\,
U'_j(x)\ve^{{N_1}-j-1} \right) \left(\sum\limits_{j=0}^k\,A_j\ve^j
\right) + \left(\sum\limits_{j=0}^N\,B_j\ve^j\right),
\end{eqnarray*}
where $A_0,\ldots,A_k,\ B_0,\ldots,B_N$ are arbitrary constants.
 
Inserting this result into (\ref{2.8}) and integrating we
come to the following Ansatz for the function $\psi(x)$:
\begin{equation}
\label{2.10}
\begin{array}{rcl}
\psi(x)&=&\left(\sum\limits_{j=-1}^{N_1-1}\,U_j(x)\ve^{N_1-j-1}\right
           )^{\frac{1}{2}} \\[4mm]
       &&\times \exp\left \{-f(\ve){\ds \int}\,{\ds {dx\over 
       \sum_{j=-1}^{N_1-1}\,U_j(x)\ve^{N_1-j-1}}} \right\},
\end{array}  
\end{equation}
where $f(\ve)=(\sum_{j=0}^N\,B_j\ve^j)(\sum_{j=0}^k\,A_j\ve^j)^{-1}$.

Substitution of the expression (\ref{2.10}) into the initial equation
after some manipulations gives the following algebraic equation:
\begin{equation}
\label{2.11}
\ve^{2N_1+1} + \sum\limits_{j=0}^{N_1-1}\,{\cal I}_j\ve^j - f(\ve)^2 = 0,
\end{equation}
where ${\cal I}_j$ are integrals of the ordinary differential equation
$U_{N_1}(x)=0$ and, consequently, are constant on the set of its
solutions.

Thus, for an arbitrary $N_1\in {\bf N}$ any solution $V=V(x)$ of the
ordinary differential equation $U_{N_1}(x)=0$ gives rise to an exactly
solvable Schr\"odinger equation. Eigenvalues of the Schr\"odinger
operator $S$ are the roots of the algebraic equation (\ref{2.11}) and
the eigenfunctions are of the form (\ref{2.10}), where $\ve$ is an
arbitrary solution of (\ref{2.11}).

It is instructive to consider in more detail the above
formulae for the case $N=N_1=1$. With this choice of $N$ the formula
(\ref{2.10}) reads as
\begin{equation} 
\label{2.12}
\psi(x)=
({2{\ve} - {{V(x)}}})^{\frac{1}{2}}\, {\exp\left\{
  {-2f({\ve})\,\left( 
    \int {dx\over {2{\ve} - {V(x)}}} \right) }\right\}},  
\end{equation}
where $f(\ve)= (A_1\ve+A_0)(B_1\ve+B_0)^{-1}$ and the function $V(x)$
is a solution of the ordinary differential equation $U_1(x)=0$, i.e.
\begin{equation}
\label{2.13}
-3{{V(x)}^2} + V''(x) = 0. 
\end{equation}

Inserting the Ansatz (\ref{2.12}) into (\ref{0.1}) yields the
following equality:
\begin{equation}
\label{2.14}
-8{{\ve}^3} + {\cal I}_0 + 8{{f({\ve})}^2} = 0, 
\end{equation}
where 
\begin{displaymath}
{\cal I}_0= {{V(x)}^3} - \frac{1}{2} {{V'(x)}^2} 
\end{displaymath}
is the first integral of equation (\ref{2.13}). Note that an
alternative derivation of the formulae (\ref{2.12})--(\ref{2.14}) has
been obtained in \cite{nik3}.

Now let us make an important remark. It is readily seen from formulae
(\ref{2.12})--(\ref{2.14}) that the function $f(\ve)$ is not obliged
to be a ratio of two first-order polynomials. It may be arbitrary. And
what is more, eigenvalues of the Schr\"odinger operator are the zeros
of the function $F(\ve) = -8{{\ve}^3} + {\cal I}_0 + 8{{f({\ve})}^2}$.
Choosing an arbitrary function $f(\ve)$ in a proper way we can get the
function $F(\ve)$ having arbitrary prescribed set of zeros. This means
that the Schr\"odinger equation (\ref{0.1}) with $V(x)$ satisfying
(\ref{2.14}) may have an arbitrary spectrum.
  
For example, if we choose $f=\sqrt{\ve^3-{\cal I}_0/8}$, then the
function (\ref{2.12}) is a solution of the Schr\"odinger equation
under arbitrary $\ve$ (the case of a continuous spectrum). Next, if we
choose $f=\sqrt{\ve^3-{\cal I}_0/8 + \sum_{j=0}^N\,A_j\ve^j}$, then a
finite discrete spectrum is obtained (eigenvalues are roots of the
$N$-th order polynomial). At last, choosing $f=\sqrt{\ve^3-{\cal
    I}_0/8 + \sin \ve}$ yields an infinite discrete spectrum
(eigenvalues are zeros of the $\sin \ve$). 

Similar results are obtained for $N=N_1=2$
\begin{eqnarray*} 
\psi(x)&=&
({8{{\ve}^2} - 4{{\ve}V(x)} + 
        {3{{V(x)}^2} - V''(x)}})^{\frac{1}{2}}\\ 
  &&\times \exp\left\{{-8f({\ve})\left( \int 
            {dx\over {8{{\ve}^2} -
                4{\ve}V(x) + 3{{V(x)}^2} - V''(x)}} 
          \right) }\right\}, 
\end{eqnarray*}
where $f(\ve)$ is an arbitrary function, $V(x)$ is a solution of the
ordinary differential equation
\begin{equation}
\label{2.15}
10{{V(x)}^3} - 5{{V'(x)}^2} - 10V(x)V''(x) + V^{(4)}(x)
=0
\end{equation}
and $\ve$ is a solution of the equation
\begin{displaymath}
-128{{\ve}^5} - 2{\ve}{\cal I}_1 + {\cal I}_0 +
128{{f({\ve})}^2} =0. 
\end{displaymath}
Here ${\cal I}_0,\ {\cal I}_1$ are integrals of equation (\ref{2.15}) of
the form
\begin{eqnarray*}
{\cal I}_0 &=& 5{{V(x)}^4} - 10V(x){{V'(x)}^2} - {{V''(x)}^2} + 
  2V'(x)V^{(3)}(x),\\
  {\cal I}_1 &=& -{{{\cal I}_0V(x)}\over 2} + {{19{{V(x)}^5}}\over 2}
  - {{{{{\cal I}_0}^2}}\over {8{{V'(x)}^2}}} - 
  {{5{\cal I}_0{{V(x)}^4}}\over {4{{V'(x)}^2}}} - 
  {{25{{V(x)}^8}}\over {8{{V'(x)}^2}}}\\
  && + 
  {{5{{V(x)}^2}{{V'(x)}^2}}\over 2} - 10{{V(x)}^3}V''(x) - 
  {{V'(x)}^2}V''(x) + {{5V(x){{V''(x)}^2}}\over 2}\\
  && + 
  {{{\cal I}_0{{V''(x)}^2}}\over {4{{V'(x)}^2}}} + 
  {{5{{V(x)}^4}{{V''(x)}^2}}\over {4{{V'(x)}^2}}} - 
  {{{{V''(x)}^4}}\over {8{{V'(x)}^2}}}.
\end{eqnarray*}

Generically, if $V(x)$ is a solution of the $2N$-th order ordinary
differential equation $U_N(x)=0$, then the corresponding Schr\"odinger
operator $S$ may have an arbitrary spectrum. Eigenvalues of $S$ are
obtained by solving the algebraic equation (\ref{2.11}) and its
eigenfunctions by substituting the corresponding values for $\ve$ into
(\ref{2.10}).

We will finish this section with one more puzzling property of the
exactly integrable models obtained. Let us denote the total
derivatives of the functions $U_j(x)$ with respect to $x$ as $W_j(x)$
and consider an infinite set of evolution equations for a function
$u=u(t,x)$
\begin{equation}
\label{2.16}
{\p u(t,x)\over \p t}=F_j[u(t,x)],\quad j>1,
\end{equation}
where the functions $F_j$ are obtained from $W_j(x)$ by formal
replacement of $V(x)$ with $u(t,x)$. Now we see that equations
obtained in this way form the famous integrable KdV hierarchy. Taking,
for example, $j=2$ yields the KdV equation
\[
{\p u(t,x)\over \p t} = {1\over 2^3}\left (6u(t,x) {\p u(t,x)\over \p
    x} - {\p^3 u(t,x)\over \p x^3}\right ).
\]

Furthermore, differentiating the relations (\ref{2.7}) we obtain the
recurrence relations determining $W_j(x)$
\[
W_{j+1}(x)= Y\, W_j(x)\equiv \left (\frac{1}{4}\p_x^2 -
V(x)-\frac{1}{2}V'(x)\p_x^{-1}\right )\, W_j(x),\quad j=0,1,\ldots
\]
with $W_0(x)=0$. The operator $Y$ above is nothing else but the
well-known recurrence operator for the KdV hierarchy \cite{ibr}. 

Next, if we formally replace $V(x)$ by $u(t,x)$ in $U_j(x)$ determined
by the recurrence relations (\ref{2.7}), then the densities of motion
constants of the KdV equation are obtained, $X$ being the recursion
operator connecting these densities.

Equations of the form (\ref{2.9}) are known in the literature as the
stationary Lax-Novikov equations. Equations of the stationary KdV
hierarchy are particular cases of equations (\ref{2.9}) with
$A_1=\cdots=A_{N-1}=0$. We have proved that any solution $V=V(x)$ of
the stationary Lax-Novikov hierarchy (\ref{2.9}) yields an exactly
solvable Schr\"odinger equation. The correspondence between solutions
of stationary KdV hierarchy and exactly solvable Schr\"odinger
equations with reflection-less potentials is known (see, e.g.\ the
paper \cite{gra} and references therein).  Moreover, it has been
established that the Schr\"odinger operators with so chosen potentials
may have an arbitrarily prescribed spectrum.  But the fact that
solutions of the stationary Lax-Novikov hierarchy (\ref{2.9}) have the
same property seems to be new.

\section*{IV.\ Conclusion}

In view of numerous excellent papers and monographs (see e.g.
\cite{ush} and the literature cited therein) devoted to
developing algebraic methods for the investigation of spectral
properties of equation (\ref{0.1}), it is, of course, not enough to
say that the problem of describing part of the spectrum of the
Schr\"odinger equation is equivalent to computing its conditional
symmetries in order to justify a necessity of introducing a new
complicated structure. Our principal motivation for looking for a
symmetry interpretation of the results obtained in this field is that
it may open a possibility
\begin{itemize}
\item{to study the spectrum of two- and three-dimensional Schr\"odinger
    equations,}
\item{to investigate \lq spectral properties\rq\ of nonlinear
    Scr\"odinger equations, and}
\item{to study spectral properties of matrix differential
    operators (say, of the Dirac operator)}
\end{itemize}
by purely algebraic means. 

For instance, there are strong evidences that a necessary condition
for a three-dimen\-si\-on\-al Schr\"odinger equation to be exactly
solvable is an invariance with respect to a three-dimensional Lie
algebra of high order symmetry operators.  We can guess that one of
the necessary conditions of \lq quasi exact solvability\rq\ of the
three-dimensional Schr\"odinger equations is a non-trivial conditional
symmetry admitted. The simplest possibility to move in this direction
is to combine the technique developed in the present paper with the
method of separation of variables \cite{mil}. In our paper \cite{zh4}
we have classified potentials $V(x_1,x_2)$ such that the corresponding
two-dimensional Schr\"odinger equation
\[
i\psi_t + \psi_{x_1x_1} + \psi_{x_2x_2} = V(x_1,x_2)\psi
\]
can be separated into three ordinary differential equations. One of
these is a first-order equation and can always be integrated by
quadratures. Two other are exactly of the form (\ref{0.1}) and can be
solved within the framework of the approach described in Sections
2 and 3.

Furthermore, the method of conditional symmetries is applicable not
only to linear partial differential differential equations but also to
nonlinear ones \cite{fu10,zh3}. Let, for example, the
one-dimensional nonlinear Schr\"odinger equation
\begin{equation}
  \label{3.1}
  \psi_{xx}=\left (\ve + V(x) + F(\psi,\psi^*,\psi_x,\psi^*_x)\right
  )\psi 
\end{equation}
be conditionally invariant with respect to an $n$-th order
Lie-B\"acklund operator 
\[
Q=\eta(x,\psi,\psi',\ldots,\psi^{(n)})\p_{\psi} + \cdots   
\]
in the sense of \cite{zh3}. Then, using a technique similar to that
developed in Section 2 we can construct an Ansatz for a function
$\psi(x)$, which gives a solution of (\ref{3.1}), provided the energy
parameter $\ve$ satisfy some algebraic equation $G(\ve)=0$. Solutions
of this equation can be interpreted as eigenvalues of the nonlinear
Schr\"odinger operator $\p_x^2-V-F$.

An interesting example is a family of nonlinear Schr\"odinger
equations suggested by {\sc Doebner} \& {\sc Goldin} \cite{doe}.
Taking the polar decomposition
\begin{equation}
\label{3.2}
\psi(t,x)={\rm e}^{r(t,x)+ i s(t,x)},
\end{equation}
and fixing the gauge $\nu_1=-1,\ \nu_2=0$ (this is always possible
\cite{nat}) we can represent the Doebner-Goldin model in the following 
way:
\begin{equation}
\label{3.3}
\begin{array}{l}
r_t + s_{xx} + 2r_xs_x=0,\\[2mm]
s_t + 2 \mu_2 r_{xx} +\mu_1 s_{xx} +4(\mu_2+\mu_5)r_x^2\\[2mm]
\quad +2(\mu_1+\mu_4)r_xs_x +\mu_3 s_x^2 +\mu_0V(x)=0,
\end{array}
\end{equation}
where $\mu_0,\ldots, \mu_5$ are model parameters.

If we impose on the solutions of (\ref{3.3}) an additional condition
$s_x=0$ (which picks out a subset of stationary solutions), then the
system obtained is consistent if and only if the conditions
\begin{equation}
\label{3.4}
r=r(x),\quad s=-\ve t,\quad \ve = {\rm const} \in {\bf R}
\end{equation}
are fulfilled.

With this choice of functions $r$ and $s$ system (\ref{3.3}) reduces
to a single equation
\begin{equation}
\label{3.5}
-\ve + 2 \mu_2 r'' + 4(\mu_2+\mu_5)(r')^2 + \mu_0V(x)=0,
\end{equation}
which is either linear ($\mu_2+\mu_5=0$) or can be linearized by the
substitution
\begin{equation}
\label{3.6}
\varphi(x)=\exp\left \{\frac{\mu_2}{2(\mu_2+\mu_5)}r(x)\right \}
\end{equation}
to become
\begin{equation}
\label{3.7}
\varphi''=\left(\frac{\ve}{4(\mu_2+\mu_5)}
-\frac{\mu_0}{4(\mu_2+\mu_5)}V(x)\right) \varphi.
\end{equation}

As established in Section 2, any equation of the form (\ref{3.7})
possesses a nontrivial high order conditional symmetry. Since the
nonlinear equation (\ref{3.5}) is equivalent to (\ref{3.7}), it
possesses high order conditional symmetry as well. Thus, the initial
Doebner-Goldin equation has a subset of solutions with non-trivial
conditional symmetry which can be effectively applied to construct
finite or even infinite (if the conditional symmetry can be reduced to
a high order Lie symmetry) set of its exact solutions.

We hope that the reasonings above are convincing enough to motivate a
further study of high order conditional symmetries of linear and
nonlinear Schr\"odinger equations in one, two and three dimensions.
It may be also very interesting to study classical and conditional
symmetries of the stationary Dirac equation in the presence of
non-vanishing electro-magnetic field and to apply these to derive a
spectrum of the Dirac operator. These problems are under investigation
now and will be a topic of our future publications.

\section*{Acknowledgments} The results of in the present paper
were obtained by the author when he was staying at the
Arnold-Sommerfeld Institute for Mathematical Physics, Technical
University of Clausthal as the Alexander von Humboldt Research Fellow.
Author wishes to thank the Director of AS Institute {\sc H.-D.
  Doebner} for hospitality. Special thanks are addressed to {\sc
  W. Fushchych} and {\sc A. Nikitin} for critical remarks and
suggestions and also to {\sc A. Ushveridze}, who introduced the
author into the exciting world of quasi exactly solvable models.
Furthermore, author express his gratitude to {\sc P. Nattermann}
for critical reading the manuscript and pointing out the possibility
of linearization of the one-dimensional Doebner-Goldin model
restricted to a subset of stationary solutions.


\begin{thebibliography}{100}
\bibitem{fu1} W.I. Fushchych, {\em Symmetry and Solutions of Nonlinear
    Mathematical Physics Equations} (Institute of
  Mathematics, Kiev, 1987), p.4.

\bibitem{fu2} W.I. Fushchych, {\rm Ukrain. Math. J.}, {\bf 39}, 116
  (1987).

\bibitem{fu3} W.I. Fushchych and A.G. Nikitin, {\em Symmetries of
    Maxwell's Equations} (Reidel, Dordrecht, 1987).

\bibitem{fu4} W.I. Fushchych, W.M. Shtelen, and N.I. Serov, {\em
    Symmetry Analysis and Exact Solutions of Nonlinear Equations of
    Mathematical Physics} (Kluwer Academic Publishers, Dordrecht,
  1993).

\bibitem{fu5} W.I. Fushchych, {\em Proceedings of the International
    Workshop ``Modern Group Analysis''} (Kluwer Academic Publishers,
  Dordrecht, 1993), p.231.  

\bibitem{cla} P. Clarkson and M. Kruskal, {\rm J. Math. Phys.}, {\bf
    30}, 2201 (1989).

\bibitem{olv1} P. Olver and Ph. Rosenau, {\rm Phys. Lett. A}, {\bf
    112}, 107 (1987).

\bibitem{win} Levi D. and Winternitz P., {\rm J. Phys. A: Math. Gen.},
  {\bf 22}, 2915 (1989).

\bibitem{fu6} W.I. Fushchych and R.Z. Zhdanov, {\rm Phys. Rep.},
  {\bf 172}, 123 (1989).

\bibitem{fu7} W.I. Fushchych and R.Z. Zhdanov, {\rm J. Math. Phys.},
  {\bf 32}, 3488 (1991).

\bibitem{fu8} W.I. Fushchych, R.Z. Zhdanov, and I.V.Revenko, {\rm
    Ukrain.  Math. J.}, {\bf 43}, 1471 (1991).

\bibitem{zh1} R.Z. Zhdanov, I.V. Revenko and W.I. Fushchych, 
  {\rm J. Math. Phys.}, {\bf 36}, 7109 (1995).

\bibitem{zh2} R.Z. Zhdanov and W.I. Fushchych, {\rm J. Phys. A: 
    Math. Gen.}, {\bf 28}, 6253 (1995).

\bibitem{blu} G.W. Bluman and J.D. Cole, {\rm J. Math. Mech.}, {\bf
    18}, 1025 (1969).

\bibitem{tur} A.V. Turbiner, {\rm Comm. Math. Phys.}, {\bf 118}, 467
  (1988).

\bibitem{shi} M.A. Shifman and A.V. Turbiner, {\rm Comm. Math. Phys.},
  {\bf 126}, 347 (1989).

\bibitem{ush} A.G. Ushveridze, {\em Quasi-exactly solvable models in
    quantum mechanics} (IOP Publ., Bristol, 1993).
 
\bibitem{olv2} P. Olver, {\em Applications of Lie Groups to
    Differential Equations} (Sprin\-ger, New York, 1986).

\bibitem{ovs} L.V. Ovsjannikov, {\em Group Properties of Differential
    Equations} (Nauka, Novosibirsk, 1962).

\bibitem{fu9} W.I. Fushchych and R.Z. Zhdanov, {\rm Ukrain. Math.
    J.}, {\bf 44}, 970 (1992).

\bibitem{fu10} W.I. Fushchych and R.Z. Zhdanov, {\rm Proc. Acad. Sci.
    Ukraine}, N5, 40 (1994).

\bibitem{zh3} R.Z.Zhdanov, {\rm J. Phys. A: Math. Gen.}, {\bf 28},
  3841 (1995).

\bibitem{fok} A.S. Fokas and Q.M. Liu, {\rm Phys. Rev. Lett.}, {\bf
    72}, 3293 (1994).

\bibitem{nik2} W.I. Fushchych and A.G. Nikitin, {\em Symmetries of
    Equations of Quan\-tum Mechanics}, (Allerton Press, New York, 1994).
  
\bibitem{nik3} A.G.Nikitin, {\rm J. Non. Math. Phys.}, {\bf 2}, 405
  (1995). 

\bibitem{kap} O.V. Kaptsov, {\rm Non. Anal., Theory, Meth. Appl.},
  {\bf 19}, 753 (1992).

\bibitem{gon} A. Gonsalez-Lopez, N. Kamran and P. Olver, {\rm Comm.
    Math. Phys.}, {\bf 153}, 117 (1993).

\bibitem{fu11} W.I. Fushchych and R.Z. Zhdanov, {\rm Sov. J. of
    Part. and Nuclei}, {\bf 19}, 1154 (1988). 

\bibitem{zh5} R.Z. Zhdanov, Preprint ASI-TPA/23/95, Technical
  University of Clausthal (1995).

\bibitem{kam} E. Kamke, {\em Differentialgleichungen L\"osungmethoden
    und L\"osungen} (Akademische Verlagsgesellschaft, Leipzig, 1976).

\bibitem{nik1} A.G. Nikitin, S.P. Onufriychuk and W.I. Fushchych,
  {\rm Theor. Math. Phys.}, {\bf 91}, 268 (1992).

\bibitem{ibr} N.Kh. Ibragimov, {\em Transformation Groups Applied to
    Mathematical Physics} (Reidel, Dordrecht, 1985).

\bibitem{gra} A.K. Grant and J.L. Rosner, {\rm J. Math. Phys.}, {\bf
    35}, 2142 (1994).

\bibitem{mil} W. Miller, {\em Symmetry and Separation of Variables}
  (Addison-Wesley, Massachusetts, 1977).

\bibitem{zh4} R.Z. Zhdanov, I.V. Revenko and W.I. Fushchych, {\rm J.
    Math. Phys.}, {\bf 36}, 5505 (1995).

\bibitem{doe} H.-D. Doebner and G.A. Goldin, {\rm J. Phys. A: Math.
    Gen.}, {\bf 27}, 1771 (1994).

\bibitem{nat} P. Nattermann, {\rm Rep. Math. Phys.}, {\bf 36}, 387
  (1995).
\end{thebibliography}
\end{document}